# Location Sensitive Embedding for Knowledge Graph Reasoning

Deepak Banerjee[1], Anjali Ishaan[2]

1) Department of Computer Science and Engineering, Calcutta University
2) Department of Applied physics, Calcutta University
{Bane.Deepak, Anjali_ishaan}@caluniv.ac.in

**Abstract**

Embedding methods transform the knowledge graph into a continuous, low-dimensional space, facilitating inference and completion tasks. Existing methods are mainly divided into two types: translational distance models and semantic matching models. A key challenge in translational distance models is their inability to effectively differentiate between 'head' and 'tail' entities in graphs. To address this problem, a novel location-sensitive embedding (LSE) method has been developed. LSE innovatively modifies the head entity using relation-specific mappings, conceptualizing relations as linear transformations rather than mere translations. The theoretical foundations of LSE, including its representational capabilities and its connections to existing models, have been thoroughly examined. A more streamlined variant, $LSE_d$, which employs a diagonal matrix for transformations to enhance practical efficiency, is also proposed. Experiments conducted on four large-scale KG datasets for link prediction show that $LSE_d$ either outperforms or is competitive with state-of-the-art related works.

## 0 Introduction

With the advent of the big data era, Google proposed knowledge graph in 2012 to improve user search, content aggregation and other services, which set off a boom in knowledge graph research [1]. Knowledge graph is a structured way to describe entities and their relationships. Entities are modeled as nodes, and relationships are modeled as different types of edges connecting the entities, which can be expressed in the form of triples $(h, r, t)$. Although knowledge graphs are well structured and defined, their potential symbolic nature poses a challenge for automatic construction and reasoning. Knowledge graph embedding simplifies relational reasoning into mathematical operations by transforming discrete entities and relationships into a continuous vector space [2]. Knowledge graph embedding methods can be broadly categorized into two main groups: translational distance models and semantic matching. Translational distance models and semantic matching models. The translational distance model models relationships as translational transformations and uses distance scores to measure the correctness of triples; the semantic matching model uses similarity scores. In addition, some studies have introduced additional information (e.g., entity types, relationship paths, time, etc.) for modeling [3-6], and these approaches are beyond the scope of this paper. The purpose of this paper is to analyze the shortcomings of the translational distance model and propose a more generalized distance model.

The constraints imposed by the existing translation distance model on triples can be summarized by a generalized formula $f_r(\boldsymbol{h}) + \boldsymbol{r} = f_r(\boldsymbol{t})$, where $f_r(.)$ is a mapping determined by relations and is usually modeled as a linear transformation. In the transE [7] model, $f_r(\cdot)$ degenerates into an identity matrix. That is to say, it requires $\boldsymbol{h} + \boldsymbol{r} = \boldsymbol{t}$. This simplification ignores the different semantics of entities under different relationships, and can't handle the relationships of "one-to-many", "many-to-one" and "many-to-many" well. In order to solve this problem, TransR [8] introduces a transformation matrix determined by the relationship between connected entities, so that $f_r(\boldsymbol{x}) = \boldsymbol{x}\boldsymbol{M}_r$; Subsequently, TransH [9] fixed the transformation matrix as a projection matrix, and the normal vector of the projection plane was determined by the relation. Models such as TransA [10] and TransD [11] introduce new constraints on the transformation matrix and greatly reduce the number of parameters.

However, these models have not been substantially improved. According to the experimental results, the performance of the carefully trained TransE model is not lower than that of the later improved model [12]. Theoretically, the translation distance model cannot model the map of the ring structure. Previous work has noticed this, but the causes of the ring structure have not been discussed from different types of relationships [13]. The following focuses on this point.

The atlas of the ring structure can be divided into three categories: (1) the ring formed by symmetrical relationship, in which the head entity and the tail entity are interchanged, and the triplet still holds; (2) The loop formed by reciprocal relationship, that is, the head/tail entities of two different relationships are the tail/head entities of another relationship; (3) The loop formed by the relationship path is the combination of different relationships. In these three cases, the phenomenon that different entities and relationships have the same representation or represent zero vectors will bring serious interference to the reasoning process, which is called "degeneration" in this paper. The root cause is that the translation distance model ignores the semantic difference between the head and the tail of the entity. Therefore, this paper proposes a position-sensitive embedding model, and it is hoped that the constraint $f_r(\boldsymbol{h}) = \boldsymbol{t}$ will be established, that is, the relationship only acts on the head entity, while retaining the original semantics of the tail entity. At the same time, the semantic transformation of the relationship to the head entity is no longer limited to translation, but a general linear transformation.

Aiming at the problem of embedding knowledge map, this paper puts forward a location-sensitive distance model. Theoretical analysis and experimental results show that this model can achieve the most advanced performance at present. In the new model, the linear time complexity can be achieved by further simplifying the transformation of relations, so that the model can be extended to large-scale knowledge map. At the same time, the probability distribution of negative sampling is improved to avoid a large number of meaningless negative samples in the model training process. Experiments show that this technique can improve the final performance of the model.

## 1 Related Work

Before introducing related work, this paper briefly introduces the symbolic system of this paper. Line vectors $\boldsymbol{h}, \boldsymbol{r}, \boldsymbol{t} \in \mathbb{R}^d$ are used to represent symbolic triplets $(h, r, t)$ (in some works, the dimensions of relations and entities can be different, so for simplicity, this paper adopts the same dimension). $s(h,r,t)$ is used to express the score of the probability of triple establishment.

Knowledge map embedding usually takes the ranking loss as the optimization objective, that is, $\min\mathcal{L} = \sum_{(h,r,t)}\sum_{(h',r,t')}[\gamma + s(h,r,t) - s(h',r,t')]_+$. Where, $[x]_+ = \max(0,x)$, $\gamma$ controls the distance between positive and negative samples. Because the knowledge map does not include negative samples, $(h',r,t')$ is obtained by randomly replacing the head entity or tail entity of the correct triplet $(h,r,t)$ (but not both at the same time). This process is called negative sampling. In the process of negative sampling, false negative triplets may be introduced, that is, the correct triplets are just unrecorded, which will be discussed later.

Cross entropy loss can also be used as the optimization objective, that is

$$\min\mathcal{L} = -\sum_{(h,r,t)} \ln p(h,r,t) - \sum_{h',r,t'} \ln\ 1 - p\ \ h',r,t'$$

In this case, the probability that the distance score is established after Sigmoid function is obtained, that is, $p(h,r,t) = \sigma(\gamma - s(h,r,t))$.

The existing research shows that [3], for the translation distance model, the ranking loss effect is better; For the semantic matching model, the effect of cross entropy loss is better. In addition, knowledge map embedding usually has restrictions on the length of entities and vectors [3], and these restrictions are cancelled in this paper.

1.1 Translational Distance Model

The translational distance model models relationships as translational transformations, where the smaller the distance between the head entity and the tail entity after translation, the greater the likelihood that the triad will be formed. From the perspective of the Unified Framework, TransE and its improved models are special cases of the distance $\|\boldsymbol{hR}_r + \boldsymbol{r} - \boldsymbol{tR}_r\|_p$ is the distance from each entity to the other. where $\boldsymbol{R}_r$ is determined by each relation; $p = 1{,}2$ is either 1-parameter or 2-parameter.

TransE [7] defines the distance score as $\|\ \boldsymbol{h} + \boldsymbol{r} - \boldsymbol{t}\ \|_p$. It can be seen that if $\boldsymbol{R}_r = \boldsymbol{I}$, it is a TransE model.

Before translational transformations, TransR [8] maps entities into different relationally determined spaces in order to capture the different semantics of entities in different relational links, i.e., $\|\boldsymbol{hR}_r + \boldsymbol{r} - \boldsymbol{tR}_r\|_p$.

TransH [9] further assumes that the mapping matrix undergoes a projection transformation, and the projection plane is determined by each relation. Let the unit normal vector of the projection plane be $\boldsymbol{w}_r$ , then the projection is $\boldsymbol{e} - \boldsymbol{e}\boldsymbol{w}_r^{\mathrm{T}}\boldsymbol{w}_r$.

TransD [11] assumes that the mapping matrix is jointly determined by the relations and their connected entities, and decomposes the matrix into the product of 2 vectors, i.e. $\boldsymbol{R}_r = \boldsymbol{w}_r^{\mathrm{T}}\boldsymbol{w}_e + \boldsymbol{I}$.

TransA [10] uses $(|\boldsymbol{h} + \boldsymbol{r} - \boldsymbol{t}|)\boldsymbol{M}_r(|\boldsymbol{h} + \boldsymbol{r} - \boldsymbol{t}|)^{\mathrm{T}}$, which assumes that $\boldsymbol{M}_r$ symmetric. The following shows that it is a special case of $\|\boldsymbol{hR}_r + \boldsymbol{r} - \boldsymbol{tR}_r\|_p$. Assuming that the mapping matrix in $\|\boldsymbol{hR}_r + \boldsymbol{r} - \boldsymbol{tR}_r\|_p$ is invertible, i.e., $\boldsymbol{R}_r^{-1}$ exists, then it can be rewritten as $\|(\boldsymbol{h} + \boldsymbol{r}' - \boldsymbol{t})\boldsymbol{R}_r\|_p$; where, $\boldsymbol{r}' = \boldsymbol{r}\boldsymbol{R}_r^{-1}$. When $p = 2$, it is $(\boldsymbol{h} + \boldsymbol{r}' - \boldsymbol{t})\boldsymbol{R}_r\boldsymbol{R}_r^{\mathrm{T}}(\boldsymbol{h} + \boldsymbol{r}' - \boldsymbol{t})^{\mathrm{T}}$. Noting that $\boldsymbol{M}_r = \boldsymbol{R}_r\boldsymbol{R}_r^{\mathrm{T}}$ is a semipositive definite matrix, we can omit the operation of taking absolute values, which is the TransA model.

In addition, TranSparse [14] considers sparse mapping matrices, and TransM [15] employs $\theta_r\|\boldsymbol{hR}_r + \boldsymbol{r} - \boldsymbol{tR}_r\|_p$ , which loosens the restriction on some of the triples by adjusting the parameters. These models are special cases.

1.2 Semantic Matching Models

These models use similarity scoring, and differ from each other in how they capture the interaction between embedded representations.

RESCAL[16] adopts quadratic scoring function, namely $\boldsymbol{hM}_r\boldsymbol{t}^{\mathrm{T}}$, to capture the interaction between potential factors of entities; DistMult [17] further simplified the matrix into diagonal matrix, and adopted $\boldsymbol{h}\mathrm{diag}\ (\boldsymbol{r})\boldsymbol{t}^{\mathrm{T}}$, which made the model unable to distinguish symmetric relations. Complex [18] studied the modeling bottleneck of point multiplication operation on antisymmetric relations, and proposed to extend Dist Mult to complex space. That is, the scoring function is Re $(\boldsymbol{h}\mathrm{diag}\ (\boldsymbol{r})\bar{\boldsymbol{t}}^{\mathrm{T}})$. Hole [19] uses the cyclic correlation operator to aggregate the interaction between the head entity and the tail entity, that is, $(\boldsymbol{h}^*\boldsymbol{t})\boldsymbol{r}^{\mathrm{T}}$; Where $[\boldsymbol{h} * \boldsymbol{t}]_i = \sum_{k=0}^{d-1}[\boldsymbol{h}]_k \cdot [\boldsymbol{t}]_{(k+i)\mathrm{mod}d}$. It can be proved that there is an equivalent hole [20] for any ComplEx model.

In addition, there is a branch of semantic matching model based on neural network, including semantic matching embedding (SME) [21], neural tensor networks (NTN) [22], etc. In this branch, the convolutional embedding model (Conve). It recombines and stacks the vectors of head entities and relations, then uses convolution layer to extract features, and finally matches with tail entities through full connection layer. Because of a large number of convolution, its computational complexity is high.

## 2 Position Sensitive Distance Modeling

In this section, we firstly analyze several types of relationships that cannot be modeled by the translational distance model; secondly, we propose an improved position-sensitive distance model, and at the same time, this paper no longer models the relationships as translations but general linear transformations, and analyzes theoretically that it has a better representational ability; then, we propose a simplified model; and lastly, we discuss the relationships and differences between the model and other existing models.

2.1 Reasons for the Failure of the Translational Distance Model

Remember the knowledge map $G = \{E, R\}$, where $E$ stands for the set of all entities, quantity is denoted as $n_e$, $R$ stands for the set of all relations, and quantity is denoted as $n_r$. This paper summarizes the relations of three specific patterns: symmetry, reciprocity and combination, and gives specific definitions.

Definition 1. If $\forall a, b \in E, (a,r,b) \to (b,r,a)$, the relation $r$ is said to be symmetrical, as shown in Figure 1a.

Definition 2. If $\forall a, b \in E, (a,r_1,b) \to (b,r_2,a)$ , the relations $r_1$ and $r_2$ are said to be reciprocal, as shown in Figure 1b。

Definition 3. If $\forall a,b,c \in E$, $(a,r_1,b) \land (b,r_2,c) \to (a,r,c)$, the relation $r$ is a combination of $r_1$ and $r_2$, as shown in Figure 1c.

As mentioned above, the circular structure brought by these three relationships leads to the degradation of the translation distance model. (1) Symmetry requires that $f_r(a) + r = f_r(b)$ and $f_r(b) + r = f_r(a)$ be established at the same time, which leads to $r = 0$; Furthermore, $f_r(.)$ is usually a reversible linear transformation, which leads to $a = b$; (2) The reciprocal relation requires that $f_{r_1}(a) + r_1 = f_{r_1}(b)$ and $f_{r_2}(a) + r_2 = f_{r_2}(b)$ be established at the same time, and there is a nontrivial solution satisfying this condition. (3) Combinatorial relations require $f_{r_1}(a) + r_1 = f_{r_1}(b), f_{r_2}(b) + r_2 = f_{r_2}(c)$ and $f_{r_3}(a) + r_3 = f_{r_3}(c)$. It will lead to $f_{r_3}(\cdot) = f_{r_2}(\cdot)$. The latter means that the same mapping function is used for different relations, which is an important reason why the improved model of TransE (including TransR, etc.) does not actually exceed that of TransE.

From the above derivation, it can be seen that the root of the problem lies in the same mapping of the head entity and the tail entity of the triple. Therefore, this paper proposes an improved distance model with position sensitivity.

### 2.2 Improved Location-Sensitive Distance Model

The most intuitive idea is that the linear transformation is reserved for the head entity of the triple, but no transformation is made for the tail entity, that is, the expected model establishes the relationship $f_r(\boldsymbol{h}) + \boldsymbol{r} = \boldsymbol{t}$ for the correct triple. Because the translation operator can be regarded as a part of the linear transformation, it can be attributed to people $f_r(\cdot)$. Finally, In this paper, a Location-Sensitive Embedding (LSE) model is proposed. The triple distance scoring function is $\|\boldsymbol{hR}_r - \boldsymbol{t}\|_p$. Similarly, the smaller the distance, the higher the possibility of the triple being established. $P$ can take 1 norm or 2 norm. Through experiments, it is better to choose 1 norm in this paper.

Three lemmas are given below, which show that LSE model can model the relationships of the above three specific patterns.

Lemma 1. Improved distance model can model symmetric relations.

Prove that according to $\boldsymbol{R}_r\boldsymbol{a} = \boldsymbol{b}, \boldsymbol{R}_r\boldsymbol{b} = \boldsymbol{a}$, it can be obtained that $\boldsymbol{R}_r\boldsymbol{R}_r = \boldsymbol{I}$. If this condition is met, there will be no additional restrictions on the embedding of the entity.

Lemma 2. The improved distance model can model reciprocal relations.

Prove that according to $\boldsymbol{R}_{r_1}\boldsymbol{a} = \boldsymbol{b}, \boldsymbol{R}_{r_2}\boldsymbol{b} = \boldsymbol{a}$, we can get $\boldsymbol{R}_{r_1}\boldsymbol{R}_{r_2} = \boldsymbol{I}$. Meeting this condition will not impose additional restrictions on the entity embedment.

Lemma 3. The improved distance model can model the combination of relationships.

According to $\boldsymbol{R}_{r_1}\boldsymbol{a} = \boldsymbol{b}, \boldsymbol{R}_{r_2}\boldsymbol{b} = \boldsymbol{c}, \boldsymbol{R}_{r_3}\boldsymbol{a} = \boldsymbol{c}$, it can be obtained that $\boldsymbol{R}_{r_2}\boldsymbol{R}_{r_1} = \boldsymbol{R}_{r_3}$, which will not be beneficial to the entity.

#### 2.2.1 Loss Function

In order to better consider the possible triplets in the unobserved samples, the cross-entropy loss with label-smoothness is experimented in this paper. For $k$ groups of samples obtained by random negative sampling, the label is not 0, but $1/k$; Where $k > 1$ is the proportion of negative sampling for each triplet. The loss function is $\min\mathcal{L} =$

$$\sum_{(h,r,t)} \left\{ \log\, p(h,r,t) - \sum_{(h',r,t')} \frac{1}{k} \log\left(1 - p(h',r,t')\right) \right\}$$

Note that this paper removes all restrictions on the norm of parameters, so the loss function has no regular term.

#### 2.2.2 Variations of the model

In the LSE model, using a matrix to represent the mapping function will cause a large amount of calculation. To solve this problem, this paper also assumes that the matrix can be replaced by diagonal matrix, that is, the mapping operation of the head entity can be carried out independently in each dimension. This model is called $\mathrm{LSE}_d$, It uses the distance scoring function $\| \boldsymbol{h}\mathrm{diag}\,(\boldsymbol{r}) - \boldsymbol{t} \|_p = \| \boldsymbol{h} \circ \boldsymbol{r} - \boldsymbol{t} \|_p$. It notes that $\| \boldsymbol{h} \circ \boldsymbol{r} - \boldsymbol{t} \|_1 = \sum_{i=1}^{d} |h_i \cdot r_i - t_i|$.

#### 2.2.3 Comparison and Connection with Other Models

Among the existing knowledge map embedding models, the translation distance model is most closely related to this model. Besides, unstructured model (um) [24] and structured embedding (SE) [25] are the same as LSE. No translation operator is used. The distance score function of UM is $\| \boldsymbol{h} - \boldsymbol{t} \|_2$, while that of SE is $\|\boldsymbol{R}_r^1\boldsymbol{h} - \boldsymbol{R}_r^2\boldsymbol{t}\|_2$. The former oversimplifies the role of the relationship, while the latter adopts similar ideas as this paper, but it introduces too many parameters, which is not conducive to the training of the model.

In addition, rescal [17] and distmult [18] can both be regarded as linear transformations of the head entities. The difference is that they use the inner product to calculate the similarity. If the vector transformation in this model is limited to the unit modulus length before and after, and $p = 2$, then $\|\boldsymbol{hR}_r - \boldsymbol{t}\|_2^2 = 2 - 2\boldsymbol{hR}_r\boldsymbol{t}^{\mathrm{T}}$. Equivalent to RESCAL. It can be seen that compared with RESCAL, this model removes the limitation of module length, so it has better representation ability. In DistMult, the head entity and the tail entity get the same score, which will lead to the model treating all relationships as symmetrical. The simplified model LSE_d proposed in this paper avoids this point.

Table 1 lists the complexity comparison results with the current advanced models. It can be seen that $\mathrm{LSE}_d$, as a linear time complexity model, can be applied to reasoning tasks of large-scale knowledge maps.

Table 1 Comparison of complexity of knowledge graph embedding models.

| Model | Space complexity | Time complexity |
|---|---|---|
| TransE[7] | $(n_r + n_e)d$ | $O(d)$ |
| TransR[8] | $(n_r + n_e)d + n_r d^2$ | $O(d^2)$ |
| DistMult[17] | $(n_r + n_e)d$ | $O(d)$ |
| HolE[19] | $(n_r + n_e)d$ | $O(d\ln\, d)$ |
| ComplEx[18] | $2(n_r + n_e)d$ | $O(d)$ |
| ConvE[23] | $(n_r + n_e)d$ | |
| $\mathrm{LSE}_d$ | $n_e d + n_r d^2$ | $O(d^2)$ |

### 2.3 Framework

Next, we briefly introduce our proposed knowledge-embedded KGNN model, partially cited in Wu etc. [20] The proposed generalized graph neural network framework for multivariate time series data consists of a graph learning module, a knowledge module, and an algorithmic modeling system. $m$ A graph convolution module (GC module), $m$ The graph learning module is composed of a temporal convolution module (TC module) and an output module. The graph learning module learns the relationships between variables and extracts such unidirectional relationships to incorporate external knowledge such as variable properties into the model. For spatial and temporal dependencies within a sequence, a hybrid jump propagation layer and an expanded starting layer are designed to capture the dependencies. However, since the automatic learning results cannot incorporate the expert knowledge in the domain, the resulting association matrix does not present the actual relationship between the variables well. In order to further improve the accuracy and generalization ability, we reduce the overfitting possibility of MTGNN by exploiting the regularization effect of the expert experience. [21,22] , Specifically, we artificially refine the correlations between variables as a knowledge matrix.

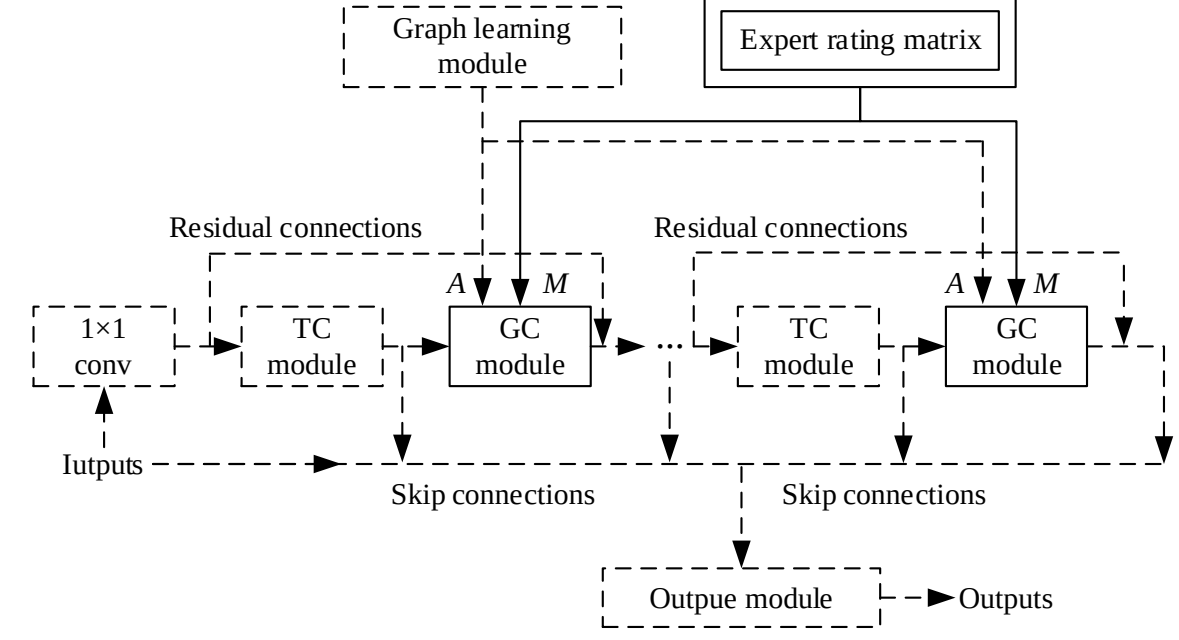

Fig.1 The framework of our model.

Its experience is utilized to score the correlations between variables. The knowledge matrix is added to the graph convolution module as a priori knowledge, and the knowledge matrix between variables and the automatically learned graph structure are combined and added to the experience mixing layer of the graph convolution module, so that the graph convolution layer is guided by the expert's experience, combined with the data-driven approach, to discover potential correlations from the data that are neglected by the expert's experience, and thus to improve the final prediction effect.

## 2.4 Prediction Methods

In this section, we introduce the prediction methods in detail, including the cited basic modeling framework [20] and the knowledge embedding method proposed in this paper. This is the background of our model for knowledge embedding and graph convolution, which consists of a graph learning module, a temporal convolution module, a jump connection layer and an output layer. We will describe in detail how to optimize the MTGNN model through external knowledge embedding by incorporating the knowledge matrix into the graph convolution module.

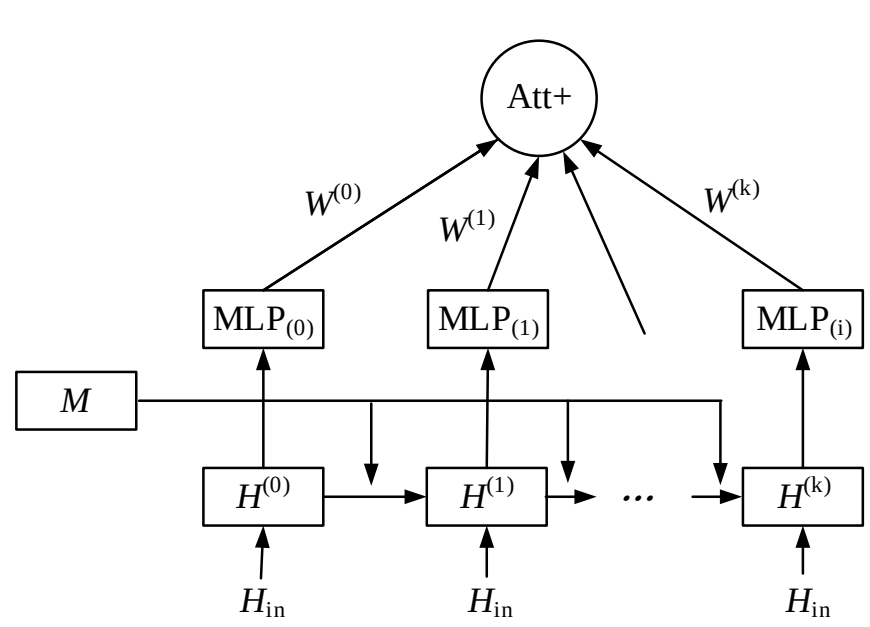

(a) Mix-hop propagation layer

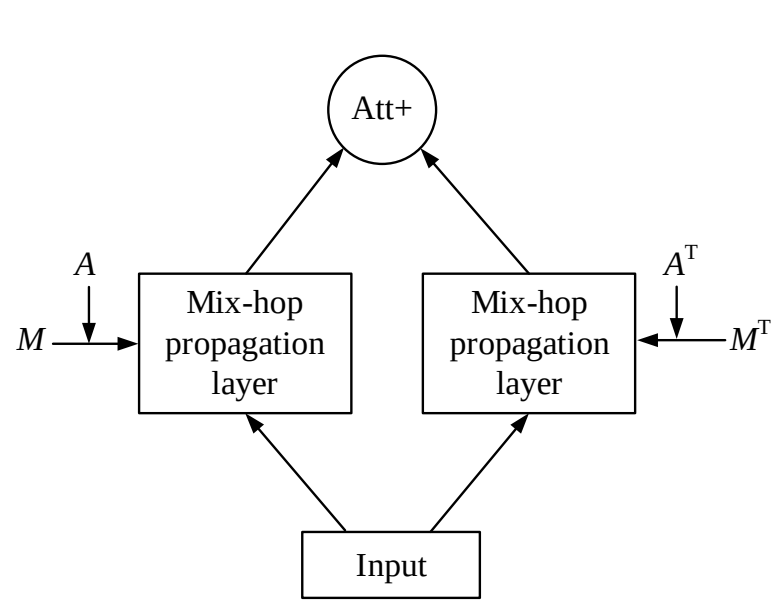

(b) Graph convolution module

Fig.2 The structure of our modules.

### 2.4.1 Graph Learning Module

The graph learning module automatically learns the graph structure in the graph adjacency matrix to obtain potential correlation information between time series data. In the prediction of multivariate time series, we hope that the change of one node condition will lead to the change of the other node condition, and avoid symmetric or bi-directional distance metrics, so we design this method to learn the unidirectional relationship.

$$M_1 = \tanh(\alpha E_1 W_1)$$
$$M_2 = \tanh(\alpha E_2 W_2)$$
$$B = \mathrm{ReLU}\left(\tanh\left(\alpha\left(M_1 M_2^{\mathrm{T}} - M_2 M_1^{\mathrm{T}}\right)\right)\right)$$

where $E_1, E_2$ denote the randomly initialized embedding matrices, which are obtained by learning during training; the $W_1, W_2$ are the linear transformation transpose matrices, which correspond to the latent features of the original data feature variables. $\alpha$ is a predefined hyperparameter, used to control $\tan h$ Saturation rate of the activation function.Eq. (4) is obtained by learning $M_1, M_2$ Obtain the correlation matrix between the original variables $B$, the association matrix is regularized using the subtraction term and the ReLU activation function if $B_{vu}$ is positive, then its diagonal counterpart $B_{vu}$ has zero value, and this equation makes the adjacency matrix asymmetric.

$B_{vu}$ The magnitude of the value indicates the strength of the correlation between the nodes, and for $B$ Each node in $B[i,:]$, we keep the first $k$ The largest value indicates connectivity, and the rest are assigned 0, From this, the adjacency matrix $A$ between nodes is obtained.

### 2.4.2 Graph Convolution Module

The main role of graph convolution module in this paper is to deal with the interdependence between nodes, and to integrate the mutual information of nodes (variables) and strongly related nodes, whose inputs come from the outputs of the previous layer of temporal convolution.In this module, the interrelationships between variables can be learned in a data-driven manner. We add a knowledge matrix to this layer to guide the graph convolution module to better learn the interrelationships between variables. The main structure of the graph convolution module is the two hybrid jump propagation layers. The hybrid jump propagation layer handles the inflow of variables at each node, and the knowledge matrix is added to the empirical mixing layer to obtain the final outflow of information by adding the outputs of the hybrid jump propagation layer and the empirical mixing layer with the attention of the two layers. The graph convolution module then fuses the outputs of the hybrid jump propagation layer through the attention layer. Fig. 2 shows the graph convolution module (GC module) and the hybrid hop propagation layer (mix-hop) for fusing knowledge matrices.

Given a graph adjacency matrix $A$, the hybrid jump propagation layer is used to process the information transfer on the neighboring feature nodes.The hybrid jump propagation layer consists of two steps: information transfer and information selection output.Given a graph structure, the information transfer step inputs the hidden node information (potential feature vector) from the previous step $H^{k-1}$ and the original state information $H_{\mathrm{in}}$, recursively passes the processed node information to the next step $H^k$.In the graph convolution module, with the number of passing layers $K$ Up to a certain number ( $K$ is a large value), the node's hidden state receives a point.Therefore, in addition to passing the processed information in the transmission process $\tilde{A}H^{(k-1)}$, while preserving a portion of the node's original state $H_{\mathrm{in}}$, so the node states passed can both preserve locality and explore deeper neighborhoods. The message passing step is defined as follows.

$$H^{(k)} = \beta H_{\mathrm{in}} + (1-\beta)\tilde{A}H^{(k-1)},$$

where, in the case of $H_{\mathrm{in}}$ is the original input state, the $\beta \in [0,1]$ is a preset hyperparameter that controls the preservation of the original state of the root node $H_{\mathrm{in}}$ and the transmitted information vector $H^{(k-1)}$ of the corresponding ratio, we empirically assign $\beta$ The initial value is set to 0.3 , and the optimization is learned gradually by back propagation. $\tilde{A} =$

$\widetilde{D}^{-1}(A+I), A$ is the adjacency matrix obtained by learning the graph structure, and $I$ is the unit matrix, the $\widetilde{D}_{ii}=1+\sum_j A_{ij}, D$ is the diagonal matrix, the $k\in[1,K]$ is the current transfer layer, the $K$ is the total number of iterations.

Instead of adding all the hidden node information directly, the information selection output layer introduces the information attention matrix $W^{(k)}$ Focusing on the information of the main hidden nodes, we filter out the useless information generated at each step and keep the most important information.In the initial state of the graph convolution module or when there is no correlation between the variables, directly summing up all the hidden node information will only output useless noise to the next layer. Where the parameter matrix $W^{(k)}$ is the attention matrix obtained by backpropagation learning in the graph convolution module. When the information of the given hidden nodes is independent of each other, it is still possible to learn the matrix by backpropagation on the $W^{(k)}$ Adjust accordingly to preserve the original node $k>0$ The information about itself is defined as follows. The information attention output step is defined as follows.

$$H_{\text{out}}=\sum_{k=0}^{K} H^{(k)}W^{(k)}$$

where $W^{(k)}$ is the first $k$ the corresponding attention matrix of the propagation layer, adjusting the correlations between features accordingly. $H_{\text{out}}$ is the output layer of the graph convolution module.

Our knowledge matrix $M$ Both are embodied in the information transfer and information selection output layers, which are used to guide our learning of the adjacency and attention matrices. In the information transfer step, a knowledge matrix is fused to the information transferred at each step, and a weighted selection of the knowledge matrix is performed as follows.

$$H^{(k)}=\beta H_{\text{in}}+(1-\beta)\left((1-\alpha_1)\tilde{A}+\alpha_1 M\right)H^{(k-1)}$$

where $\alpha_1\in[0,1]$ is a preset ratio parameter that regulates the graph learning matrix and knowledge matrix, and its initial value is set to 0.1, which is gradually learned and optimized through backpropagation.

As a result, we improve the learning efficiency and effectiveness of the graph convolution module by inputting the experience matrix, so that the convolution layer, guided by domain knowledge and experts' experience, and combined with the data-driven approach, can discover the neglected potential relationships between variables from the data, and thus improve the final prediction effect of the model.This approach is similar to the original loss equation (1) with the additional knowledge matrix used for training the regularization model. The effect is equivalent to making the association matrix obtained from the self-learning of the graph network $A$ To the knowledge matrix $M$ Close.

2.4.3 Temporal Convolution Module

The temporal convolution module extracts high-dimensional temporal features through a set of standardized dilated 1D convolution filters. The module consists of two dilated 1D inception layers [20].A hole initial layer is connected to a tangent hyperbolic activation function as a filter, and another hole initial layer is connected to a sigmoid activation function, both of which are used to control the amount of information output to the next module.

When determining the appropriate filter size, the Wu et al. [20]] proposed a time-initiation layer consisting of four filter filters, each of which is of size $1\times 2, 1\times 3, 1\times 6$ and $1\times 7$, covering all possible time signal periods with a combination of 4 filter sizes; when dealing with very long sequences, the Wu et al. use null convolution to avoid the use of very deep networks or very large filters, thus reducing the model complexity. For example, let the expansion factor of each layer be exponential in $q(q>1)$ .The initial expansion factor is assumed to be 1, then a convolutional kernel of size $c$ The $\varepsilon$ The corresponding receptive field size of the layer null convolutional network is

$$R=1+\frac{(c-1)(q^{\varepsilon}-1)}{q-1}$$

The hole initial layer combines the start and hole convolution, and the input of the time convolution part is denoted as $z$, given a feature b, the input expression of 1D sequence $z_{,,b}\in R^{l}$. The input to the 1st time-convolution module comes from $1\times 1$ The output of the convolution, and the rest of the inputs to the temporal convolution module come from the residual connections and the output of the convolution of the previous layer of the graph, as shown in Fig. 2.The output from the $f_{1\times 2}\in\mathbb{R}^2, f_{1\times 3}\in\mathbb{R}^3, f_{1\times 6}\in\mathbb{R}^6, f_{1\times 7}\in\mathbb{R}^7$ comprise the set of filters obtained through learning, and thus the null initial layer is denoted as

$$z_{:,b}^{l}=\operatorname{concat}\left(z_{:,b}^{l-1}\star f_{1\times 2}, z_{:,b}^{l-1}\star f_{1\times 3}, z_{:,b}^{l-1}\star f_{1\times 6}, z_{:,b}^{l-1}\star f_{1\times 7}\right)$$

where $z_{:,b}^{l}$ is the current $l$ The input of the layer, the $f$ is the filter, $\star\star$ is the convolution symbol, concat is the concatenation operator.

The outputs of the 4 filters are truncated to the same length according to the largest filter and concatenated in the channel dimension, the null convolution is expressed as

$$\left(z_{:,b}^{l}\star f_{1\times g}\right)[j]=\sum_{i=0}^{g-1} f_{1\times g}(i)\star z_{,,b}^{l}(j-e\times i)$$

where $e$ is the null factor, the $g$ Indicates the size of the filter, the $l$ refers to the first $l$ Layer.

2.4.4 Jump Connection Layer and Output Module

The jump connection layer is $1\times L_i$ Standard convolution.

$$\left(h\star f_{1\times L_i}\right)[j]=\sum_{k=0}^{L_i-1} f_{1\times L_i}(k)\star h(j-k)$$

where $h$ denotes the input of the hopping connection layer, the $L_i$ is the input $X$ to the first $i$ The length of the hopping connection layer, thus regularizing the dimension of the output vector of each temporal convolutional layer, so that each output has the same sequence length.

The output module converts the dimensionality of the input data into the desired output dimensions, and consists of two $1\times 1$ A standard convolutional layer is formed. When the output is expected to be the predicted value at a certain point in time, the output dimension is 1 dimension. When the output is expected to be continuous $n$ When the predicted values of the time points, the output dimension is $n$ Dimension.

## 3 Experimental Results and Discussion

In this section, we firstly introduce the dataset, the link prediction task of the knowledge graph, and the evaluation metrics used in the experiments, and then discuss the influence of hyperparameters and the selection techniques, and finally list the performance of the improved distance model and compare it with the current best performance model.

It is worth mentioning that research work has shown that many knowledge graph embedding methods (including TransE, DistMult, etc.) can exceed their original published results after fine tuning. This is one of the criticisms of the improvement of the flat distance model, i.e., the original comparison methods are not fine-tuned, but in fact, some of the baseline methods still have much room for training improvement. In order to ensure the fairness of the comparison, we have chosen [12,23,26,27] as the source of the comparison.

## 3.1 Experimental Setup

### 3.1.1 Data sets

The model of this paper is evaluated on four public datasets, and their statistical characteristics are shown in Table 2.

Table 2: Number of entities, relationships, triples and cuts in the datasets

| Data set | *nr* | *ne* | Number of training sets | Number of validation sets | Number of test sets |
|---|---|---|---|---|---|
| FB15k | 1 345 | 14 951 | 483 142 | 50 000 | 59 071 |
| FB15k237 | 237 | 14 541 | 272 115 | 17 535 | 20 466 |
| WN18 | 18 | 40 943 | 141 442 | 5 000 | 5 000 |
| WN18RR | 11 | 40 943 | 86 835 | 3 034 | 3 134 |

(1) FB15k vs. FB15k237. FB15k is a subset of the fact database Freebase [28]. The main drawback of this database is that 81% of the triples in the test set can be derived from inverse relations. These relations have been removed in FB15k-237, thus making it more challenging.

(2) WN18 and WN18RR. WN18 is a subset of the word hierarchy database WordNet [29]. Compared with Freebase, WordNet has a larger number of entities. Compared with Freebase, WordNet has more entities, fewer relationship types, and fewer triples, which indicates that the interconnections of each entity node are sparse on average, and this poses a major problem for knowledge graph representation learning. WN18 also has many reversible relationships, which are removed by WN18RR.

### 3.1.2 Link Prediction and Evaluation Metrics

Link prediction is essentially an ordering problem. In the testing stage, for each triad, all the entities in the graph are replaced with the head or tail entities, and then all the replacement results are ranked in ascending order by their distance scores. The ranking results are evaluated using two types of indicators.

(1) mean reciprocal rank, MRR). Average the reciprocal of the ranking of the correct entity among all entities, that is, $\sum_{i=1}^{N}(1/\text{rank}_i)/N$. The higher the ranking of the correct entity, the higher the MRR.

(2) Top $n$ hit rate (Hits@n). The proportion of correct entities in the top $n$ entities. This paper makes statistics when n = 1, 3 and 10 respectively.

Some papers also use mean rank as an evaluation index, but it is too affected by extreme samples and is not as stable and objective as MRR. In addition, the triad generated after substitution is not necessarily a negative sample, and it may also exist in the knowledge graph. The above calculation may underestimate the performance of the model. In order to get a fair evaluation, a "filtering" setting is usually used, i.e., the correct triples are filtered out from the ranking results before the evaluation metrics are calculated.

## 3.2 Experimental Environment and Hyperparameters

The code in this paper is based on the PyTorch framework, and is trained on a single Titan X GPU (12 GB).

The hyperparameters involved in the LSE and $\text{LSE}_d$ models proposed in this paper include three categories: (1) model hyperparameter, embedding dimension $d \in \{100{,}200$, interval between positive and negative samples $\gamma \in \{1,2,\cdots,30\}$. (2) optimization algorithm superparameter, and model training adopts randomness. SGD) algorithm, which involves hyperparameters and its selection range is learning rate $l \in \{5\times10^{-4}, 10^{-4}, 5\times10^{-5}, 10^{-5}\}$, and the batch data size is $b \in \{128,256,512,1024\}$,. (3) Negative sampling hyperparameter. In this paper, according to transh [9] method, the head entity or tail entity is randomly replaced by Bernoulli distribution to avoid attracting too many false negative triples in the process of uniform sampling. For each triplet, the ratio of random negative sampling is $g \in \{128,256,512,1024\}$.

Through the selection of verification set, this paper uses $d = 500$ for all four data sets. For FB15k, $\gamma = 24, l = 5\times10^{-4}, b = 256, p = 10^5, g = 256$; For FB15K-237, $\gamma = 9, l = 5\times10^{-4}, b = 1024, p = 10^5, g = 256$; For Wn18,$\gamma = 12, l = 5\times10^{-4}, b = 512, p = 5\times10^4, g = 1024$; For Wn18RR, $\gamma = 6, l = 5\times10^{-4}, b = 512, p = 10^5, g = 1024$. In the process of parameter adjustment, it is found that expanding the batch data size and increasing the negative sampling ratio will improve the performance when the experimental memory is allowed.

## 3.3 Experimental Results and Analysis

Tables 3 and 4 compare the performance of the LSE, LSEd models with the other models on the link prediction task.

Tables 3 and 4 compare the performance of LSE, LSE _d model and other comparison models in link prediction tasks, respectively. It can be seen that the models proposed in this paper have achieved the best performance on FB15k-237 and WN18RR data sets. On FB15k and WN18 data sets, the performance of the model proposed in this paper is basically the same as that of the best model. In addition, $LSE_d$ has a significant performance improvement compared with LSE, which shows that fewer parameters are more suitable for training, and also shows that the assumption that each dimension is independent is true.

For each dataset, the following findings are observed.

Table 3 Link prediction performance of the improved distance model on the FB15k and WN18 datasets.

| Model: FB15k | FB15k | | | | WN18 | | | |
|---|---|---|---|---|---|---|---|---|
| | MRR | Hits@1 | Hits@3 | Hits@10 | MRR | Hits@1 | Hits@3 | Hits@10 |
| TransE[12] | 0.463 | 0.578 | 0.578 | 0.578 | 0.495 | 0.113 | 0.888 | 0.943 |
| DistMult[26] | **0.798** | | | **0.893** | 0.797 | | | 0.946 |
| HolE[19] | 0.524 | 0.402 | 0.613 | 0.739 | 0.938 | 0.930 | 0.945 | 0.949 |
| ComplEx[18] | 0.692 | 0.599 | 0.759 | 0.840 | 0.941 | **0.936** | 0.945 | 0.947 |
| ConvE[23] | 0.745 | 0.670 | **0.801** | 0.873 | **0.943** | 0.935 | **0.946** | **0.956** |
| LSE | 0.701 | 0.643 | 0.733 | 0.795 | 0.873 | 0.842 | 0.898 | 0.926 |
| LSEd | 0.742 | **0.680** | 0.783 | 0.848 | 0.929 | 0.910 | 0.944 | 0.955 |

Notes. Bold indicates the best performance.

Table 4 Link prediction performance of the improved distance model on the FB15k237 and WN18RR datasets

| Model | FB15k237 | | | | WN18RR | | | |
|---|---|---|---|---|---|---|---|---|
| | MRR | Hits@1 | Hits@3 | Hits@10 | MRR | Hits@1 | Hits@3 | Hits@10 |
| TransE[12] | 0.294 | | | 0.465 | 0.226 | | | 0.501 |
| DistMult[26] | 0.241 | 0.155 | 0.263 | 0.419 | 0.430 | 0.390 | 0.440 | 0.490 |
| ComplEx[18] | 0.247 | 0.158 | 0.275 | 0.428 | 0.440 | **0.410** | 0.460 | 0.510 |
| ConvE[23] | 0.325 | **0.237** | 0.356 | 0.501 | 0.430 | 0.430 | 0.440 | 0.520 |
| LSE | 0.299 | 0.189 | 0.312 | 0.470 | 0.421 | 0.378 | 0.450 | 0.494 |
| LSEd | **0.331** | **0.237** | **0.366** | **0.522** | **0.450** | 0.407 | **0.465** | **0.537** |

Notes. Bold indicates the best performance.

(1) DistMult achieves the highest performance on the FB15k dataset. This suggests that symmetric relationships are the dominant relationship model in this dataset. In fact, about 81% of the test set triples can be directly inferred from the symmetry of the relations [4]. It can be seen that the results of DistMult on other datasets are not as good as that of the model proposed in this paper, so it can be said that the model proposed in this paper is more general than DistMult.

(2) In the WN18 dataset, the main relationship patterns are symmetric and inverse; in the WN18RR dataset, the main relationship patterns are symmetric (e.g., relationship also see, similar to, etc.). On both datasets, the performance of TransE decreases significantly, which indicates that it cannot symmetrize relations well. This is consistent with the theoretical analysis in Section 3.

(3) On FB15k-237 and WN18RR data sets, the combination of relationships also occupies a very important part. In FB15k-237, the effect of TransE is not much different from other advanced algorithms, which is consistent with the theoretical analysis in Section 3. Similarly, the model proposed in this paper can also model the combination of relationships well.

Taken together, the experimental results in this paper correspond well with the theoretical analysis, which fully demonstrates that the models LSE and $LSE_d$ have good knowledge graph representation learning ability.

## 4 Conclusion

In this paper, a location-sensitive model for knowledge graph embedding representation is proposed, and the effectiveness of the proposed improved model is verified theoretically and experimentally, and the superior performance is achieved on the task of link prediction in large-scale knowledge graphs.

This paper notes that there are still some problems that need to be further studied. (1) In the combinatorial relationship, there is a special case that needs to be considered separately, that is, the combination formed by the same relationship. This model usually exists in the knowledge map of professional fields with more hierarchical structures. For example, (forearm, partOf, limbs) and (limbs, partOf, human body) coexist (forearm, partOf, (2) The representation of knowledge is uncertain. In KG2E [30] model, entities and relationships are no longer modeled as a point in space, but a Gaussian distribution. The application of position-sensitive embedding proposed in this paper needs to be further improved.